\documentclass[aps,twocolumn,preprintnumbers,amsmath,amssymb,showpacs]{revtex4}
\usepackage{graphicx}\usepackage{epsfig}
\begin{document}

\title{Anomalous thermodynamics and phase transitions of neutron-star matter}
\author{C. Ducoin$^{1}$}
\author{K.H.O. Hasnaoui$^{2}$}
\author{P. Napolitani$^{1}$}
\author{Ph. Chomaz$^{2}$}
\author{F. Gulminelli$^{(1)}$ \footnote{member of the Institut Universitaire de France}}

\affiliation{$^{1}$~LPC (IN2P3-CNRS/Ensicaen et Universit\'e), F-14076 Caen c\'edex, France}
\affiliation{$^{2}$~GANIL (DSM-CEA/IN2P3-CNRS), Blvd. H. Becquerel, F-14076 Caen c\'edex, France}

\begin{abstract}
The presence of the long-range 
Coulomb force in dense stellar matter implies that the total charge
cannot be associated with a chemical potential, even if it is a 
conserved quantity.
As a consequence, the analytical properties of the 
partition sum are modified, changing the order of the phase 
transitions and affecting the possible occurrence of critical 
behaviours.
The peculiar thermodynamic properties of this system can be 
understood introducing a model hamiltonian where each charge is 
independently neutralized by a uniform background of opposite 
charge.
Some consequences on the characteristics of mixed-phase structures 
in neutron star crusts and supernova cores are discussed.
\end{abstract}
  

 
\maketitle

\section{Introduction}

	Among the major quests in investigating the properties of
  	compact-star matter, is the underlying equation of state.
	In particular, exotic phases (the so-called pasta phases) are predicted  
	in the outer crust of neutron stars as well as in the inner core 
	of exploding supernovae. In this latter physical situation matter 
	attains temperatures of several MeV
	\cite{bethe,lattimer,lattimer2}, 
	and the possible existence of 
	such dishomogeneous structures is known to have some relevance for hydrodynamic 
	properties of stellar matter and neutrino transport
	\cite{watanabe,horowitz}.


	The interplay between the short-range nuclear force and the long-range 
	non-saturating Coulomb interaction is responsible for the existence
	of such phases. First calculations of the composition of the stellar crust 
	were done using the Liquid Drop Model
	or in the Thomas-Fermi approximation 
	\cite{pasta_first0,pasta_first1,pasta_first2,pasta_first3,pasta_first4}
	with an a-priori assumption on the different species constituting the matter 
	\cite{pasta_first0,pasta_first4}, or on the shape of the pasta structures 
	\cite{pasta_first1,pasta_first2,pasta_first3} : the preferred shape was
	essentially determined by the balance between surface and Coulomb energies.
	In these studies, as well as in more recent ones along the same lines
	\cite{pethick,shen,haensel,watanabe_ldm}, charge neutrality is imposed only 
	globally while charge fluctuations are allowed on any (small) scale, 
	meaning that the effect of the electron background is neglected. 
	Then 
	the problem reduces 
	\cite{ohnishi} to a simple application of Gibbs 
	phase-coexistence conditions in multi-fluid systems 
	\cite{glendenning}. As a consequence, 
	the transition from the homogeneous outer core to the pasta phase(s)  is systematically 
	considered as first order in all these works.

However, the conditions to define a thermodynamic limit
in systems involving non-saturating forces 
are not trivial and phase transitions may have 
specific properties analogous to those discussed for finite 
systems 
\cite{phase_trans}.
 In this paper we show that the suppression of one thermodynamic degree of freedom
due to the divergence of the Coulomb energy-density for any net charge at the thermodynamic limit implies the disappearance of phase transitions or a modification of their order. Specifically, 
 we show that the crust-core transition is not first order but continuous.
 
Another limitation of the first seminal works on the phase structure of stellar
matter is that most calculations were done in the mean-field approximation, which is 
known to be especially poor in describing phase transitions. Recent calculations with 
semiclassical models beyond the mean field
\cite{watanabe,horowitz} 
naturally include thermal fluctuations and allow the rearrangement of the proton
distribution due to the Coulomb field under the explicit constraint of charge 
neutrality. The inclusion of such fluctuations was expected to lead to
an increased matter opacity to neutrino scattering with important consequences
on the supernova cooling dynamics
\cite{horowitz,margueron}. 
Such a coherent neutrino-matter scattering is not only expected 
at low temperature, where the pasta-core transition 
was supposed to be first order, but even more in the case of the occurrence 
of a critical point in the post-bounce supernova dynamics with the associated
phenomenon of critical opalescence
\cite{margueron,wata_rev}.
Surprisingly, the expected increase in the static form factor 
is not observed in finite-temperature molecular-dynamics calculations
\cite{horowitz2}, 
indicating a fluctuation suppression respect to the simplistic scenario
of a first-order phase transition.
 
On the basis of simple scaling arguments, we show in this paper that critical 
  behaviors can survive up to the thermodynamic limit
  only if the particle-density fluctuations associated 
  with the mixed pasta phases do not correspond, close to the critical point,
  to charge-density fluctuations. 
  In turn, this would imply a strong increase with temperature of free-proton drip
  or, alternatively, a very strong polarization of the electron field, leading 
  to a complete charge screening of the pasta structures.
  Since both effects are likely to be unphysical, critical fluctuations 
  and the associated critical opalescence 
  are expected to be quenched. 

\section{Thermodynamic features of (proto-)neutron stars}

\subsection{Thermodynamics of charged systems}

	Let us consider the dense matter in neutron-star 
crusts and supernova cores formed of electrons ($e$), neutrons 
($n$) and protons ($p$). 
	The microscopic Hamiltonian reads
\begin{equation}
	\hat{H}=\hat{H}_{np}+\hat{K}_{e}+\hat{V}_{ee}+\hat{V}_{pp}+\hat{V}_{ep}
	\,\, ,
	\label{EQ:H}
\end{equation}
where $\hat{H}_{np}$ is the nuclear strong interaction including 
the nucleon kinetic energy, $\hat{K}_{e}$ is the electron kinetic 
term and $\hat{V}_{ii^{\prime }}$
is the Coulomb interaction between different types of particles ($i \in \{e,n,p\}$) :
\begin{equation}
	\hat{V}_{ii^{\prime }} = 
	\frac{\alpha q_{i}q_{i^{\prime }}}{1+\delta_{ii^{\prime }}}
	\int \frac{\rho _{i}(\vec{r})\rho _{i^{\prime}}
	(\vec{r}^{\prime })}
	{\left| \vec{r}-\vec{r}^{\prime }\right|}
	\mathrm{d}\vec{r}\mathrm{d}\vec{r}^{\prime } 
	\,\, ,
\end{equation}
where 
$\alpha$ is the fine-structure constant and
$\rho _{i}(\vec{r})$ is the local density of the particle 
of type $i$ and charge $q_{i}=\pm 1$.

	Since the Coulomb field is a long-range interaction, the 
existence of a thermodynamics, i.e. the convergence of a 
thermodynamic limit and the equivalence between statistical 
ensembles, is not guaranteed
\cite{phase_trans}.
	Let us first consider the canonical ensemble with densities 
$\rho _{i} = N_{i}/\Omega$, where $\Omega$ is the
volume containing a number $N_i$ of particles $i = n$, $p$ or $e$. 
	If the net charge density $\rho _{c}=\rho _{p}-\rho _{e}$ 
is not strictly zero at the thermodynamic limit, the Coulomb energy 
per unit volume $\langle V_{c} \rangle/\Omega\propto \rho _{c}^{2}\Omega^{2/3}$ diverges. 
The strict charge neutrality 
demanded by the thermodynamic limit guarantees the additivity of 
thermodynamic potentials as well as ensemble equivalence.  
	Indeed, the monopole contribution to 
the Coulomb energy between two separated neutral systems 
is identically zero.
The longer-range multipole interaction 
is the dipole-dipole term $E\propto \vec{D}_{1}\cdot\vec{D}_{2}/R^{3}$,
where $\vec{D}$ is the dipole moment in each subsystem.
This moment is proportional to $ N_{c}R$, where 
$N_{c}$ is the localized charge number 
and $R$ the maximum size of the actual dipole. 
To avoid the divergence of the Coulomb-energy density, 
$N_{c}$ must scale at maximum proportionally to $R^{2},$ indicating 
the convergence of the dipole part 
of the inter-system energy density, and 
the additivity of the two subsystems at the thermodynamic limit.
This reasoning can be extended to all multipole-multipole interactions.
The Coulomb interaction between neutral systems 
then behaves like short-range interactions 
\cite{phase_trans},
leading to ensemble equivalence 
according to the standard demonstration of the Van Hove 
theorem.   

\subsection{Thermodynamic consequences of charge neutrality}

It is important to remark that the strict neutrality constraint 
discussed above is 
very different from the 
trivial condition of global charge neutrality 
of the stellar object, which is universally recognized. 
We have just argued that neutrality is a necessary thermodynamic condition: 
this means that charge dishomogeneities can appear only 
at the microscopic level and no locally charged domain can exist 
at a macroscopic scale, even if this scale is small compared to the size of 
the star. From a practical point of view,
this means that in hydrodynamical calculations the condition 
$\rho_p(\vec{r})=\rho_e(\vec{r})$ 
has to be imposed at every location $\vec{r}$, and that this neutrality constraint 
must be explicitly applied at the scale of the Wigner Seitz 
cell in neutron-star crusts. This can have important effects 
on the phase structure of matter even if the electron background is 
assumed to be uniform, as we develop in great detail in the next sections. 

	The important consequence of the charge-neutrality constraint
is that the canonical free-energy 
density $f$ is defined only for $\rho _{c} = 0$.
	Hence $f\left( T,\rho_{n},\rho_{p},
\rho_{e} \right) = f\left( T,\rho_{n},\rho \right) $ 
with $\rho =(\rho_{p}+\rho_{e})/2$ 
and the chemical potential associated with $\rho_c$ can not be defined
since the free energy is not differentiable in the total-charge 
direction.
	We can notice that this can also be deduced in a grand-canonical treatment.  
Indeed the divergence of the Coulomb-energy density for 
$\rho_c\neq 0$
forces the grand-canonical partitions to fulfill  
the constraint $\rho _{p}=\rho _{e}$ exactly for all couples of chemical
potentials $\mu_{e}$ and $\mu _{p}$. 
	If the chemical potentials associated with 
$\rho_c$ and $\rho$ are introduced in the form
$\mu _{c} = (\mu _{p}-\mu _{e})/2$  and 
$\mu =\mu _{e}+\mu _{p}$ respectively, 
the grand potential per unit volume $g$
results independent of $\mu _{c}$, so that $g(T,\mu _{n},\mu ,\mu _{c}) = 
g(T,\mu _{n},\mu )$.
Indeed, densities are grand-potential derivatives : the condition on the total charge 
$\rho _{c}=0$ then corresponds to a flat grand-potential in the chemical-potential direction $\mu_{c}$.

	This suppression of one degree of freedom 
	arises from the thermodynamic limit and 
should not be confused with an additional constraint
as it is often done in the literature, where the two conditions
of charge neutrality (expressed as $n_p=n_e$) 
and $\beta$ equilibrium (expressed as $\mu_p=\mu_e$)
are treated on the same footing
\cite{ohnishi} or 
confused
\cite{haensel}. 
	Additional constraints, such as constant particle 
fraction or chemical ($\beta$) equilibrium, { may or may not be 
realized in the supernova evolution;} they are restrictions of the 
accessible states and do not affect the thermodynamical properties, 
which are state functions. 
Conversely, charge neutrality has to be fulfilled for each (macroscopic) physical state.
This changes the the number of degrees of freedom of the thermodynamic potentials, 
which directly affects the thermodynamics. 
In particular, phase coexistence can occur only between two neutral phases. 
The practical consequences of this thermodynamic requirement are detailed in the next sections.

\subsection{Quenching of mean-field instabilities in stellar matter}

%
%
\begin{figure}[tbh!]
\begin{center}
\includegraphics[angle=0, width=0.9\columnwidth]{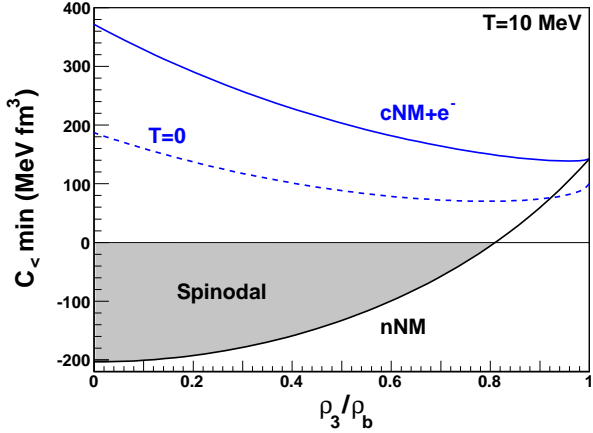}
\caption{(Color online)
Minimal free-energy curvature of homogeneous matter 
at the fixed finite temperature $T=10$ MeV 
in the mean-field approximation, obtained with the Sly230a interaction.
Without Coulomb interaction : nNM (neutral Nuclear Matter).
Switching on Coulomb interaction : cNM+e (charged Nuclear Matter neutralized by an electron gas).
Dotted line : nuclear matter with electrons at T = 0.
 }
\label{fig:1}
\end{center}
\end{figure}
	These considerations 
are especially relevant when phase transitions are concerned.
	Phase transitions should 
be analyzed by
looking at the derivative and curvature properties of 
$f$ in the three-dimensional state-variable space $(T,\rho_{n},\rho).$ 
	In particular, a first-order phase transition 
at a given temperature 
identifies with the
linear behavior of the free-energy density $f$ between at least 
two points $A=(\rho _{n}^{A},\rho ^{A})$ and 
$B=(\rho_{n}^{B},\rho^{B})$, meaning that
all intensive parameters in $A$ and $B$ are equal
\cite{glendenning}. 
	However, since the chemical potential associated with the 
total charge is not a thermodynamic variable, 
the condition $\mu^{A}=\mu^{B}$ does not imply 
as it is often assumed in the literature
\cite{pethick,lattimer,ohnishi,kubis} that the 
chemical potentials $\mu_{e}$ and $\mu_{p}$ are both identical in the two phases
($\mu_{e}^{A}=\mu_{e}^{B}$ and $\mu_{p}^{A}=\mu_{p}^{B}$).
	The difference in the chemical potential of charged particles
is counterbalanced by the Coulomb force:
	as some electrons move from one phase to the other 
driven by the chemical-potential difference, the Coulomb force 
reacts forbidding a macroscopic charge to appear. 	
	To illustrate this point, we consider a mean-field 
approximation, where the free energy of the system is defined as the 
sum of independent baryon $f_{b}$ and electron $f_{e}$ free 
energies,
\begin{equation}
	f(T,\rho_{n},\rho ) = 
	f_{b}(T,\rho _{n},\rho_p=\rho)
	+f_{e}(T,\rho_e=\rho )
	\,\, . \label{free_energy}
\end{equation}
	To spot a phase transition we can study the convexity 
anomalies of $f$ looking at the curvature matrix 
\begin{equation}
	C =
	\left( 
	\begin{array}{ll}
	  \partial \mu_{n} / \partial \rho_{n} 
	& \quad\quad
	  \partial \mu_{n} / \partial \rho_{p} 
	\\ 
	  \partial \mu_{p} / \partial \rho_{n} 
	& 
	  \partial \mu_{p} / \partial \rho_{p} 
	+ \partial \mu_{e} / \partial \rho_{e}
	\end{array}
	\right) 
	\,\, , \label{curvature}
\end{equation}
where we have introduced the chemical potentials $\mu_{n} =$  
$\partial f_{b}/\partial \rho_{n}$ , $\mu_{p} =$ $\partial f_{b}/ 
\partial \rho_{p}$ and $\mu_{e} =$ 
$\partial f_{e}/\partial \rho_{e}$. 
	The additional term $\chi_{e}^{-1} =
\partial \mu_{e} / \partial \rho_{e}$ in the matrix 
modifies the stability conditions with respect to the nuclear-matter 
part, i.e. to the curvature of $f_{b}$. 
	In general, we can expect
a quenching of the instability: since the electron susceptibility $\chi_{e}^{-1}$
is always positive, 
the instability conditions corresponding to 
$\mathrm{Det}\,C\leq 0$ or $\mathrm{tr}\,C\leq 0$ are more 
difficult to fulfill. 
	Moreover, due to their very small mass, electrons are 
almost always a relativistic degenerate Fermi gas:
\begin{equation}
 \chi_e^{-1}
= \hbar c (\frac{\pi^2}{9 \rho_e^2})^{1/3} \label{e_pres}
\,\, .
\end{equation}
Since $\chi_{e}^{-1}$ is large at subsaturation densities, 
the quenching is expected to be strong.
A quantitative application is shown in Fig.1, which displays the minimal eigenvalue of the free-energy curvature matrix eq.(\ref{curvature}) 
as a function of the isospin asymmetry $\rho_3/\rho_b=(\rho_n-\rho_p)/(\rho_n+\rho_p)$, evaluated 
with the effective Sly230a Skyrme interaction
\cite{camille3,Chabanat}.

At the thermodynamic limit, the nuclear free-energy density appearing 
in eq.(\ref{free_energy}) is defined for each temperature and each point 
$(\rho_n,\rho_p=\rho)$ in the density plane by a Legendre transform of the Gibbs free-energy density
\begin{equation}
f_b=g+ \sum_{q=n,p} \mu _{q}\rho _{q}
\,\, .
\end{equation}
In the mean-field approximation, 
this latter can be approximated using independent-particle averages only 
\cite{Vautherin,camille1} :
\begin{equation}
g\simeq - T \ln z_{0}- \frac{1}{\Omega}\left( 
\langle\hat{W}\rangle-\langle\hat H \rangle\right)
\,\, .
\end{equation}
In this expression, 
$z_0$ is the grand-canonical partition sum of a non-interacting Fermi gas of free protons and neutrons with effective mass $m^*_q$, and chemical potentials $\mu^*_q$ shifted of the value of the mean field $\mu^*_q=\mu_q-U_q$.
$\langle\hat{W}\rangle=\sum_q \langle \hat{p}^2/2m^*_q +U_q \rangle$ is the average single particle energy, $\langle\hat H \rangle=\Omega\cal{H}$ is the average energy
with the chosen Skyrme functional, and the local mean field and effective mass are defined as partial derivatives of the energy density $\cal{H}$ with respect to the 
particle densities $\rho_q$ and the momentum densities $\tau_q=\langle p^2\rangle /\hbar^2$ :
\begin{equation} 
U_q=\frac{\partial \cal{H}}{\partial \rho_q} \;\; ; \;\;
\frac{\hbar^2}{2 m^*_q}=\frac{\partial \cal{H}}{\partial \tau_q} \,\, .
\end{equation}
All averages are thermal averages over the modified single-particle Fermi distribution
\begin{equation}
n_{q}(p)=\frac{1}{1+exp(\beta (p^{2}/2m_{q}^{*}-\mu _{q}^{*}))} \,\, .
\label{EQ:distribution}
\end{equation}

From Figure 1 we can see that ordinary neutral nuclear matter (nNM) shows a wide unstable (spinodal) region characterized by a direction of negative convexity of the free energy.
In this region, any density fluctuation of homogeneous matter is spontaneously 
amplified, and leads to phase separation. 
This result is drastically modified 
considering charged nuclear matter neutralized by an electron gas (cNM+e) :
the free energy is everywhere convex, and this stays
true at zero temperature (dashed line). Since the extension of the spinodal region monotonically decreases with increasing temperatures, this latter result implies that stellar matter does not present
any thermodynamic instability 
in this model. This quenching is due to the fact that matter dishomogeneities 
cause charge dishomogeneities, which cannot be fully screened by the electron gas because of its high incompressibility, and are thus forbidden at any (macroscopic) scale. As a consequence, the different structures in the stellar crust (free nucleons and nuclei, bubbles or pasta) have to be considered as intrinsically microscopic objects, and cannot be treated as different phases in coexistence, obeying standard phase equilibrium rules, as it is often done in the literature
\cite{pasta_first4,shen}. 
We will see in section \ref{subsec:pasta-ext} 
that such microscopic charge fluctuations naturally appear beyond the mean field, 
and they indeed dominate the density region between the liquid core and the gazeous outer crust.
Such structures are however associated to an important energetic cost, 
which, as we will demonstrate, leads to an expansion of the dishomogeneous phase.    

If the qualitative effect of electron quenching of the instability is independent of the model, the net quantitative effect on the spinodal zone depends on the parameters of the nuclear interaction: the different Skyrme forces we have analyzed
\cite{camille3} all lead to a complete instability suppression, while a small spinodal region still subsists
at zero temperature in NL3 RMF stellar matter calculations 
\cite{constanca}.
%

\subsection{Suppression of first-order transitions in stellar matter\label{first_order}}

In order to expose how the charge-neutrality constraint 
affects the thermodynamics of star matter,
we now study the core-crust transition in two different approaches.
One corresponds to the physical situation, 
for which electro-neutrality has to be ensured by the correlation 
between proton and electron density distributions. 
For the other approach, we consider a model system
where each charged particle species is independently  
neutralized by a uniform background of opposite charge. 

The modified Hamiltonian reads 
\begin{equation}
	\hat{H}^{\prime } = \hat{H}_{np}+\hat{K}_{e}
	+\hat{V}_{ee}^{\prime}+\hat{V}_{pp}^{\prime }
	+\hat{V}_{ep}^{\prime }  
	\,\, ,
	\label{EQ:Hprime}
\end{equation}
where the interaction with the opposite  background charge has been 
subtracted to the various Coulomb terms
\begin{equation}
	\hat{V}_{ii^{\prime }}^{\prime } = 
	\frac{\alpha q_{i}q_{i^{\prime}}}{1+\delta_{ii^{\prime }}}
	\int \frac{\delta \rho_{i}(\vec{r})\delta 
	\rho_{i^{\prime }}(\vec{r}^{\prime })}
	{\left|\vec{r}-\vec{r}^{\prime}\right| }
	\mathrm{d} \vec{r}\mathrm{d} \vec{r}^{\prime }  
	\,\, ,
	\label{EQ:fluct}
\end{equation}
with $\delta \rho _{i}(\vec{r}) =\rho_{i}(\vec{r})-\langle\rho_{i}\rangle .$
This system does not present any divergence so that 
the free energy $f^{\prime }(T,\rho _{n},\rho_{p},\rho _{e})$ 
can always be defined.
When $\rho_{p}=\rho_{e}$ the introduced backgrounds 
cancel out exactly. In this case,
the canonical ensemble of our initial stellar problem is equivalent 
to the canonical ensemble of the modified Coulomb system along the 
$\rho _{p}=\rho _{e}$ surface : 
\begin{equation}
	f(T,\rho_{n},\rho ) = 
	f^{\prime} (T, \rho_{n}, \rho_{p}=\rho,  \rho_{e}=\rho)
	\,\, .
	\label{EQ:f-fprime}
\end{equation}
The two pressures are connected by the relation
\begin{equation}
	g(T,\mu_{n},\mu)=g^{\prime}(T,\mu_{n}^{\prime},
	\mu_{p}^{\prime},\mu_{e}^{\prime })  
	\,\, ,
	\label{EQ:g}
\end{equation}
where the chemical potentials 
$\mu _{i}^{\prime} = \partial f^{\prime }/\partial \rho_{i}$ 
of the different species $i=n,$ $p,$ $e,$ are linked 
to the physical chemical potentials 
by the relations $\mu_{n}=\mu_{n}^{\prime }$, 
$\mu =\mu_{p}^{\prime }+\mu_{e}^{\prime }$. 
Let us however notice that this equivalence breaks down in phase coexistence,
since a phase mixture in the neutralized model 
belongs to the physical model only if $\rho_c=0$ in each phase separately.

In the neutralized Coulomb system, all chemical potentials and densities are allowed.
This model may present phase transitions according to the 
usual phenomenology of multifluid systems
\cite{glendenning}.
In particular, first-order phase transitions are 
characterized by the equality of all intensive parameters between 
two points $A^{\prime }$ and $B^{\prime }$ of the particle-density space: 
$(\rho_{n}^{A^{\prime}},\rho_{p}^{A^{\prime}},\rho_{e}^{A^{\prime}})$ 
and 
$(\rho_{n}^{B^{\prime}},\rho_{p}^{B^{\prime}},\rho_{e}^{B^{\prime}})$. 
The linear behavior of $f^{\prime}(T,\rho _{n},\rho _{p},\rho _{e})$ 
between $A^{\prime }$ and $B^{\prime }$ corresponds to 
a single point in the intensive-parameter space 
$(T,\mu _{n}^{\prime },\mu _{p}^{\prime },\mu _{e}^{\prime })$
with different densities (i.e. pressure derivatives) on both sides. 

An illustration of this approach is given by figures 2 and 3,
presenting mean-field calculations of such neutralized model,
where nuclear and electron contributions are treated independently.
The strong hamiltonian is parametrized by the Skyrme Sly230a interaction
\cite{Chabanat}
and the nuclear-matter thermodynamics is evaluated within the mean-field approximation, 
as in the previous section 
\cite{camille1,camille2,camille3}. 
Until a critical temperature, nuclear matter (nNM) 
presents a first-order liquid-gas phase transition.
Electrons, conversely, are treated as a single homogeneous Fermi gas. 
This results in a violation of the relation $\rho_p=\rho_e$ in each phase, since
we have $\rho_p>(<)\rho_e$ in the dense (dilute) phase. 
In other words, $\rho_c=\rho_p-\rho_e$ is an order parameter of the phase coexistence.
This is permitted within the neutralized model
where proton and electron charge are independently cancelled in each phase
according to eq.(\ref{EQ:fluct}), suppressing the Coulomb contribution.

%
\begin{figure}[tbh!]
\begin{center}
\includegraphics[angle=0, width=0.9\columnwidth]{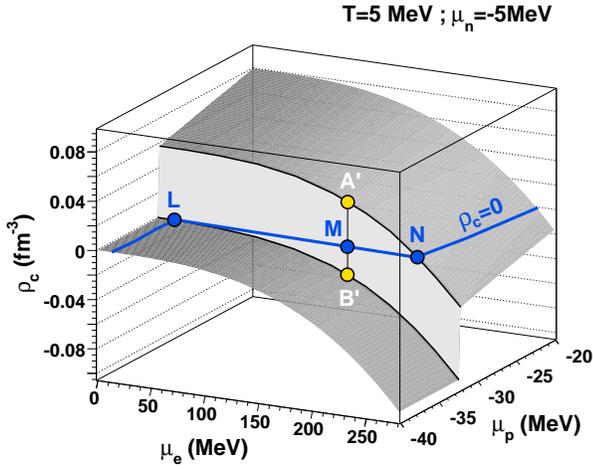}
\caption{
(Color online)
Net charge density $\rho_c=-\partial g'/\partial\mu'_c$ as a function of the proton and electron chemical potential for a neutralized Coulomb system at fixed temperature and neutron chemical potential, with the Sly230a effective interaction.
 The solid line gives the physical constraint $\rho_c=\mathrm{cte}=0$. 
The neutralized model presenting a first order phase transition with a  
$\rho_c$  discontinuity, the physical constraint forces the system 
to follow the coexistence line, leading to a continuous transition for the physical system. 
}
\label{fig:2}
\end{center}
\end{figure}
%
%
\begin{figure}[tbh!]
\begin{center}
\includegraphics[angle=0, width=0.9\columnwidth]{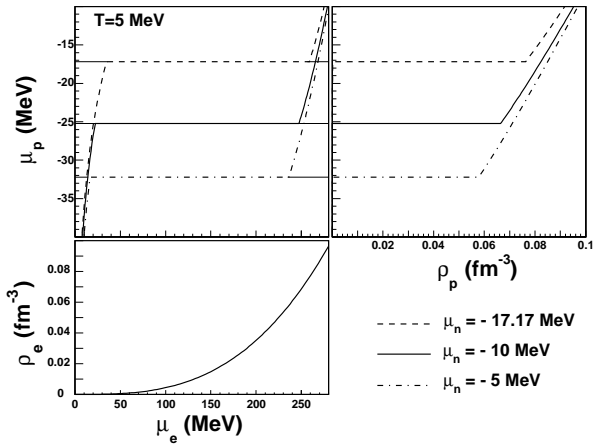}
\caption{
Upper left : representation of the constraint $\rho_p=\rho_e$ 
as paths in the chemical-potential plane ($\mu_e,\mu_p$).
Upper right : for each given neutron chemical potential, proton density is determined by $\mu_p$.
Lower part : electron density is uniquely determined by $\mu_e$, according to eq.(\ref{e_pres}).
$\mu_n=-17.17$ MeV corresponds to the coexistence zone of symmetric matter, higher  $\mu_n$ gives a matter increasingly neutron rich.
}
\label{fig:3}
\end{center}
\end{figure}

Figure 2 emphasizes the role of $\rho_c$ as an order parameter
for a phase coexistence occurring in the neutralized model.
In the neutron-star phenomenology, this neutralized system
corresponds to nuclear-matter calculations
\cite{pethick} which do not explicitly
impose the neutrality constraint on the scale of the Wigner-Seitz cell.
The high-density phase represents the homogeneous liquid core, 
while the low-density one is the unbound vapor.
Let us consider a system such that $\rho_c=0$. 
The corresponding path is represented on Figure 3 as a projection
in the chemical-potential plane ($\mu_e$,$\mu_p$), 
for different values of the neutron chemical potential $\mu_n$.
Starting from a high-density homogeneous system,
and decreasing the density,
the liquid-gas phase coexistence is reached at point $L$.
Afterwards, the global requirement $\rho_c=0$ has to be fulfilled by phase coexistence
between two phases for which $\rho_c \neq 0$. 
An  exemple is given by points $A^{\prime}$ and $B^{\prime}$, 
which obey the phase-equilibrium condition of equal intensive parameters.
This is a first-order phase transition.
Similar features are found for any fixed under-critical temperature.

However, this does not correspond to the physical situation,
where electro-neutrality imposes $\rho_c=0$ at any macroscopic scale.
Let us consider the line $\rho_c=0$ represented on Figure 2. 
In the neutralized model, this path crosses a coexistence region (N-L portion).
Turning to the physical system, the states of this region can not be described 
in terms of phase coexistence, as discussed previously : 
they instead consist in microscopic fluctuations 
constituting a single dishomogeneous phase.
In this restrained $\rho_c=0$ subspace, 
the grand-potential derivatives remains continuous
(\emph{e.g.} proton density evolves continuously) :
the physical free energy does not present a linear behavior,
which means that the phase transition is no more of first order.
However, the path $\rho_c=0$ is non-analytical when
it enters the region of $\rho_c$ discontinuity
(points $L$ and $N$ on Figure 2).
The pressure is non-analytical at such points,
which trace continuous transitions in the physical system.

To summarize, 
as the neutralized-model system enters the coexistence region 
and undergoes a first order phase transition, 
the physical system restrained to the subspace $\rho_c=0$ 
turns into a dishomogeneous phase \emph{via} a continuous transition. 
The only way to keep a first-order phase transition in the physical system would be 
to have no $\rho_{c}$ discontinuity at the phase transition in the neutralized system,
which could be realized only if the electron distribution followed 
closely the proton distribution leading to a complete screening.

In the neutron-star phenomenology, the mixed phase consists in
a Coulomb lattice of neutron-rich nuclei at zero temperature, 
or the pasta phase(s) at finite temperature. 
Most calculations presented in literature treat the  
electron dynamics at the mean-field level 
either by considering electrons as a uniform background
\cite{pethick,watanabe,magierski}, 
or allowing for a slight polarization of the electron gas, 
leading to a screening length of some tens of fm
\cite{horowitz}. 
An explicit calculation of the electron screening in the RMF coupled
to the electric field
\cite{maruyama} 
confirms this assumption.
In this condition, the clusters (nuclei or pasta) 
show a net positive charge at their (microscopic) scale, 
meaning that the total charge is an order parameter
for the transition in the model neutralized system.
As a result, the physical crust matter cannot be described as a phase coexistence :
the core-crust transition occurs between uniform matter and a mixed phase 
presenting microscopic dishomogeneities.
This transition is necessarily continuous, contrary to what is almost universally
assumed in the literature
\cite{pasta_first4,pethick,haensel,watanabe_ldm,wata_phasediag}.
Of course, for such dishomogeneous structures to emerge,
calculations beyond the mean field are necessary, 
as will be discussed in the next section.
In this case the phase diagram can be more complicated 
and present other bifurcations if the different pasta structures, 
here represented by a single coexistence region,
correspond to different phases as suggested by molecular-dynamics calculations
\cite{wata_phasediag}.
The limited increase of the static form factor to neutrino scattering
observed in the numerical calculations of 
ref.\cite{horowitz2} 
is consistent with the thermodynamic arguments developed in this section.

\subsection{Quenching of critical behavior in stellar matter}

%
%
\begin{figure}[tbh!]
\begin{center}
\includegraphics[angle=0, width=0.9\columnwidth]{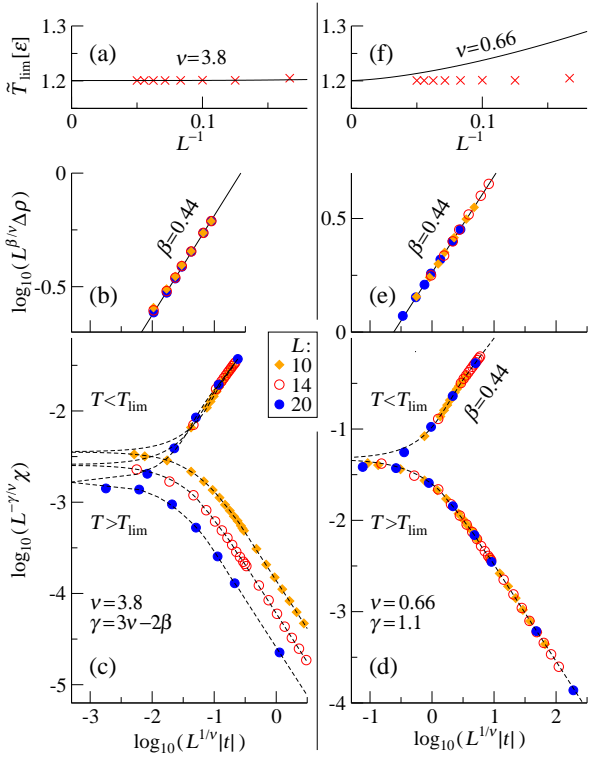}
\caption{(Color online)
Finite-size scaling for the neutralized Ising model with long-range interactions. 
Upper part : limiting temperature as a function of the lattice linear size giving the exponent $\nu$. Medium part : scaling of the order parameter giving the consistence between $\beta$ and $\nu$. 
Lower part : scaling of the susceptibility giving the consistence between $\gamma$ and $\nu$. 
The right and left parts of the figure give two different ways of fixing the critical exponents. 
None of them gives a coherent scaling of all the quantities.
 }
\label{fig:4}
\end{center}
\end{figure}
Turning now to the mixed-phase phenomenology of high-temperature supernova cores,
it is interesting to remark that
long-range Coulomb interactions also strongly affect critical behaviors.  
Indeed, the Coulomb-energy density can be expressed as a 
function of charge fluctuations as 
\begin{equation}
	\frac{\langle\hat{V}_{c}^{\prime }\rangle}{\Omega}=\frac{\alpha }{2\Omega}\int 
	\frac{\sigma_c (\vec{r},\vec{r}^{\prime })}
	{\left| \vec{r}-\vec{r}^{\prime }\right|} 
	\mathrm{d} \vec{r}\mathrm{d} \vec{r}^{\prime }=2\pi \alpha \int 
	\sigma_{c}(r) r \mathrm{d} r 
	\,\, ,
	\label{crit}
\end{equation}
where 
$\sigma (\vec{r},\vec{r}^{\prime }) = \langle\delta \rho_{c}(\vec{r})\delta \rho_{c}(\vec{r}^{\prime })\rangle$ 
is the charge-density fluctuation, and translational and  
rotational invariance imposes $\sigma _{c}(\vec{r},
\vec{r}^{\prime }) = \sigma _{c}
(\left| \vec{r}-\vec{r}^{\prime}\right| )$.   
	If the electron screening is a small effect even at high 
	temperature, i.e. if $\rho_c$ keeps being an order parameter 
	up to the critical point, then
$\sigma _{c}(r)$ is expected to scale as 
\begin{equation}
	\sigma _{c}(r)\propto r^{2-d-\eta }  
	\,\, ,
	\label{EQ:scaling}
\end{equation}
where $d$ is the space dimensionality and $\eta$ is a critical
exponent which turns out to be close to zero in most physical 
systems ($\eta=0.017$ for the liquid-gas universality class).
	A critical point would then correspond to a divergent 
Coulomb energy. 
	Hence, only three behaviors are allowed.
	Either the critical temperature goes to infinity, or the 
order parameter becomes independent of 
the total charge or
 the phase transition in the neutralized system ends	
at a first-order point. The corresponding second-order point 
in the physical system cannot then be a critical point,
 and critical opalescence is suppressed.
Concerning supernova cores, the presence of critical opalescence  
may have important consequences on the opacity to neutrino 
scattering
\cite{margueron}, 
and neutrino transport is crucial 
for simulation studies of supernova explosions 
\cite{horowitz,watanabe}.
According to this qualitative argument, we do not expect 
a strong increase of matter opacity to neutrino scattering.
To have some quantitative estimation, it is clear that 
calculations of stellar phase structure going beyond the 
mean-field approximation are necessary.

Our recent finite-size-scaling calculations 
within a Lattice Gas model 
confirm the suppression of critical behavior in stellar
matter, as we now show\cite{paolo}.
In this schematic but exactly solvable model, 
each site of a three-dimensional lattice of $\Omega=L^3$ sites  
is characterized by an occupation number $n_i=0,1$ and an effective
charge $q_i=n_i-\sum_{j=1}^\Omega n_j/\Omega$. The effective charge represents the 
proton distribution screened by a uniform electron 
background.
 
The schematic Hamiltonian $H=H_N+H_C$ with
\begin{eqnarray}
H_N &=& -\frac{\epsilon}{2}\sum_{<ij>}n_i n_j \\ 
H_C &=&   \frac{\kappa}{2}\sum_{i\neq j} \frac{q_i q_j}{r_{ij}}
         =\frac{\kappa}{2}\sum_{i\neq j} n_i n_j C_{ij}  \label{ising}
\end{eqnarray}
is introduced to study  the interplay 
of nuclear-like ($H_N$) and Coulomb-like forces ($H_C$).
	$\sum_{<ij>}$  is a sum extended over closest
neighbors, and $r_{ij}$ is the distance between sites $i$ and $j$.
	The short-range and long-range interactions are characterized
by the coupling constants $\epsilon$
and $\kappa = \alpha \rho_0^{1/3} x^2$ respectively, where $\rho_0$
is the nuclear saturation density, and $x$ is the proton fraction.
 
To accelerate thermodynamic convergence, the finite lattice is repeated
in all three directions of space a large number $N_R$ of times. The Coulomb interaction with the different replicas of each site is analytically calculated and shown to be equivalent to a renormalization of the long-range couplings $C_{ij}$ 
\cite{paolo}.
The phase diagram of the model is evaluated with standard Metropolis techniques
\cite{paolo} 
and shown to contain for all simulated lattice sizes $L$ a coexistence region terminating at a limiting temperature $\tilde{T}_{lim}(L)$.
	The quenching of criticality can be formally verified in terms of 
critical exponents.
	If the asymptotic value of the limiting temperature 
$T_{lim}=\lim_{L\rightarrow\infty} \tilde{T}_{lim}(L)$ 
did correspond to a critical point, any generic thermodynamic observable $Y$ characterized by the critical exponent $\Phi$ should fulfil close to 
$T_{lim}$:
\begin{equation}
Y\left ( L,t \right )=f\left (L/\xi(t) \right ) | t |^{\Phi},
\label{finite_size_scaling}
\end{equation}
with $t = T/T_{\mathrm{lim}}-1$.
Scale independence close to the critical point imposes for the scaling function
$f(s)\propto s^{-\Phi/\nu}$, where $\nu$ rules the divergence of the correlation length, 
$\xi\propto t^{-\nu}$.
Specifying eq.(\ref{finite_size_scaling}) to the behavior of the limiting temperature gives 
\begin{equation}
|\tilde{T}_{lim}(L) -T_{lim}| \propto L^{-1/\nu}.\label{scale1}
\end{equation} 
The scaling of the order parameter
$\Delta\rho=\langle n\rangle_{L} - \langle n\rangle_{G}$ and of the
susceptibility $\chi = \sum^{\Omega}_{ij} \langle \delta n_i \delta n_j \rangle /T $
results
\begin{eqnarray}
L^{\beta/\nu}\Delta\rho &\propto& \left ( L^{1/\nu}|t| \right ) ^\beta \;\;\; T<\tilde{T}_{lim}(L), L \gg \xi \label{scale2} \\
L^{-\gamma/\nu}\chi &\propto& \left ( L^{1/\nu}|t| \right ) ^{-\gamma} \;\;\; T \geq \tilde{T}_{lim}(L), L \gg \xi \label{scale3}
\end{eqnarray}
while both expressions should become independent of $L$ and $T$ 
as soon as  the critical point is approached for each given finite size, 
$L\approx \xi$.
For $T<\tilde{T}_{lim}(L)$, $\chi$ contains also
jumps between the low-density solution $\langle n \rangle_{G}$ 
and the high-density one $\langle n \rangle_{L}$,
and should obey first-order scaling as $t^{-2\beta}\chi \propto L^d $.
Introducing the hyperscaling relation
$d = (\gamma+2\beta)/\nu$, we get 
\begin{equation}
L^{-\gamma/\nu}\chi \propto \left ( L^{1/\nu}|t| \right ) ^{2\beta} \;\;\; T<\tilde{T}_{lim}(L), L \gg \xi \label{scale4}
\end{equation}

Fig.4 illustrates that finite-size scaling is violated for the neutralized long-range Ising model at the approach of the limiting temperature: it is not possible to find a thermodynamically consistent set of critical exponents fulfilling at the same time eqs.(\ref{scale1},\ref{scale2},\ref{scale3},\ref{scale4}).
If we fix $\beta$ and $\nu$ from eqs.(\ref{scale1},\ref{scale2}) 
(panels a and b in Fig.4), the value of $\gamma$ is fixed from the hyperscaling relation.
The calculations with different lattice sizes do not collapse on a single curve 
in the representation of eqs.(\ref{scale3}) (Fig.4c), 
even if the first order scaling eqs.(\ref{scale4}) is nicely respected
up to $\tilde{T}_{lim}(L)$. If conversely we fix $\gamma$ and $\nu$ such as 
to fulfil eq.(\ref{scale3}) (Fig.4d), this does not affect the behavior of the order parameter eq.(\ref{scale2}) (Fig.4e), but the scaling of the limiting temperature eq.(\ref{scale1}) is violated (Fig.4f). 

	The loss of critical behaviour has been already observed in Ising models  
with long-range frustrating interactions, where the coexistence  
region was seen to end at a first-order point
\cite{Tarjus}.
This effect was absent in the mean-field approximation and was
attributed to fluctuations.   
   
 The fact that no enhancement is observed in the static form 
factor of clusterized matter in the molecular-dynamics 
simulations by Horowitz and coworkers
\cite{horowitz2}
may also be an indication of this fluctuation quenching.

\subsection{Extension of the pasta phases \label{subsec:pasta-ext}}

%
%
\begin{figure}[tbh!]
\begin{center}
\includegraphics[angle=0, width=0.9\columnwidth]{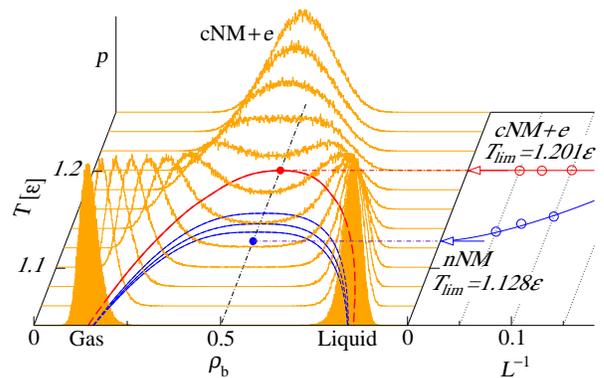}
\caption{(Color online)
Metropolis calculations of the density distributions of the neutralized Ising model with long range interactions at different temperatures, for a cubic lattice of linear size $L=10$ and with
a proton fraction $\rho_p/\rho_b=0.3$ ($cNM$). The extension of the coexistence zone, corrected for finite size effects through finite size scaling, is also shown. The lines labelled $nNM$ give the
coexistence zone of neutral matter for different lattice sizes, and the extrapolation to the thermodynamic limit.
 }
\label{fig:5}
\end{center}
\end{figure}
	Another remarkable effect that should manifest beyond the 
mean-field approximation is an extension of the 
mixed-phase pasta region in stellar matter with respect to standard 
uncharged nuclear matter.
	Even if at the ending point of the coexistence region the 
correlation length may not diverge as discussed in the previous section, 
we can still 
expect that it will be characterized by large correlated structures.
	As we infer from eq.(\ref{EQ:fluct}) and eq.(\ref{crit}), 
in the neutralized system the Coulomb energy will be minimized in 
homogeneous partitions corresponding to pure phases, and it will be
maximal in clusterized partitions.
This increase of the energy difference between the pure phases and the mixed phase
results in an extension of the coexistence region, 
\emph{i.e.} an extension of the dishomogeneous phase corresponding to the mixed partitions
situated in between the pure phases for each temperature.
Consequently, the limiting temperature for the existence of such mixed phase 
is increased with respect to the uncharged system.
This effect has been recently observed 
in the neutralized long-range Lattice-gas model presented in the previous section
\cite{paolo}, as shown in Fig.5.
The phase diagram of this model (cNM line in Fig.5) is compared to the 
one obtained when the Coulomb interaction is neglected, $\kappa=0$ in eq.(\ref{ising}) (nNM lines in Fig.5). For both models the phase diagram is extracted from the Metropolis simulated density
distribution in the grand-canonical ensemble, also shown in the figure for different chosen temperatures.

As we have argued in Section \ref{first_order}, at the mean-field level
the two models only differ for the presence of the electron pressure in cNM,
which is not added in Fig.5 and would transform the coexistence curve into a second-order transition line. For a given value of $\rho_b$, the charged-particle density $\rho_p=\rho_e=x\rho_b$ is fixed and so is the electron pressure (see eq.(\ref{e_pres})),
meaning that the microstates explored by the calculation with (cNM) or without (nNM) Coulomb in the canonical ensemble are the same at the level of mean field.
Fig.5 shows that this is not true once the phase diagram is numerically calculated from the exact Hamiltonian without any mean-field approximation.
The limiting temperature decreases with the lattice size for nNM fulfilling finite-size scaling with standard 3d-Ising exponents
\cite{paolo}, 
while it is almost completely independent of $L$ when Coulomb is accounted for, reflecting the finite correlation lenght at the limiting point and criticality quenching 
discussed in the previous section. More important, the limiting temperature 
is increased for increasing strenght of the Coulomb field (about 6\% for the proton fraction $x=1/3$ considered in Fig.5).
This expansion of the coexistence region is due to the energy cost 
of fragmented inhomogeneous configurations at $\rho/\rho_0\approx 0.5$ respect
to the the largely uniform proton-charge distribution in pure phases, whose Coulomb energy is almost exactly compensated by 
the electron background. This effect is entirely due to the screening effect of the electrons, under the assumption that they can be approximated by a fixed background. 

As such, the coexistence-region expansion is specific of the stellar problem, while other physical systems subject to Coulomb frustration 
 exhibit the opposite behavior.
This is notably the case of frustrated Ising ferromagnets, 
as well as of finite atomic nuclei.
	In such cases the Coulomb repulsion is known to reduce the 
limiting temperature
\cite{Bonche1985,Lee2001,Tarjus,Raduta2002,Gulminelli2003}.
	This reduction is also a usual expectation in the 
astrophysical context
\cite{lattimer,pasta_first1,pasta_first3,
pethick,haensel,watanabe_ldm,glendenning}.
 An amplification of the mixed-phase region at zero temperature
has already been reported in 
ref.\cite{maruyama}. In this work 
a consistent treatment of the proton-density rearrangement effect 
due to the Coulomb field under the constraint of strict charge neutrality
in the Wigner Seitz cell is seen to de-favour, at each given total density, 
the formation of large charged structures 
(rods respect to droplets, bubbles respect to tubes). The net effect is 
a density amplification of the mixed phase.
Our results imply that this effect should persist 
at finite temperature.
An increase of the maximum temperature for the 
coexistence region might be expected, and the mixed-phase 
phenomenology may be relevant for the proto-neutron-star structure 
in a wider temperature range than usually expected
\cite{glendenning}. 

\section{Conclusion}

	In this paper we have shown that long-range Coulomb forces 
in dense stellar matter require a non-trivial discussion on the 
definition of the thermodynamic limit.
	When the constraint of total charge is not 
strictly zero, the divergence of the Coulomb energy implies that
the chemical potential associated with the total charge 
loses its thermodynamic meaning.
	Hence, the grand potential, i.e. the system pressure, 
and its derivatives, i.e. the particle densities, result independent 
of the charge chemical potential.
	The suppression of a thermodynamic degree of freedom
strongly affects the thermodynamics of 
the charged system and the phase-transition phenomenology. 
In general, the Coulomb field 
modifies the non-analytical properties of the partition sums and 
changes the order of the transitions: the first-order core-crust 
transition obtained when the coulomb energy is disregarded, turns into 
a continuous transition from the homogeneous core phase 
to a mixed phase (the so called pasta phases), 
when the coulomb energy is accounted for. 
	A critical point can be found at finite temperature if and
only if the net charge does not present fluctuations.
	Only in this case, critical opalescence in stellar matter 
can be compatible with the long-range Coulomb force.


\end{document}